**Explainable convolutional neural network model provides an alternative genome-wide association perspective on mutations in SARS-CoV-2**

Parisa Hatami[1], Richard Annan[2], Luis Urias Miranda[3], Jane Gorman[4], Mengjun Xie [1], Letu Qingge[2], Hong Qin[1,5*]
1 Department of Computer Science and Engineering, University of Tennessee at Chattanooga, Chattanooga, TN, U.S.A
2. Department of Computer Science, North Carolina Agricultural and Technical State University, Greensboro, NC, USA
3. Department of Mathematics, Shenandoah University, VA
4. Department of Biology, Catholic University of America, D.C, U.S.A.
5. School of Data Science, Department of Computer Science, Old Dominion University, VA. U.S.A.
* Corresponding author, hqin@odu.edu

# Abstract

Identifying mutations of SARS-CoV-2 strains associated with their phenotypic changes is critical for pandemic prediction and prevention. We compared an explainable convolutional neural network (CNN) approach and the traditional genome-wide association study (GWAS) on the mutations associated with WHO labels of SARS-CoV-2, a proxy for virulence phenotypes. We trained a CNN classification model that can predict genomic sequences into Variants of Concern (VOCs) and then applied Shapley Additive explanations (SHAP) model to identify mutations that are important for the correct predictions. For comparison, we performed traditional GWAS to identify mutations associated with VOCs. Comparison of the two approaches shows that the explainable neural network approach can more effectively reveal known nucleotide substitutions associated with VOCs, such as those in the spike gene regions. Our results suggest that explainable neural networks for genomic sequences offer a promising alternative to the traditional genome wide analysis approaches.

Deep learning and machine learning have emerged as potent tools for connecting genotypes to phenotypes [1–4]. A significant challenge in applying machine learning models in biology is their lack of interpretability, known as the 'black box' problem[2]. This challenge could potentially be addressed by the explainable artificial intelligence (xAI) approach[5]. The SHapley Additive exPlanations (SHAP)[6] framework is an xAI tool designed to dissect and quantify the impact of each feature on the prediction outcomes.

Classic methods for GWAS often rely on linear (or logistic) regression models that assume independence and linearity in the contributions of genetic loci to phenotypic variation. However, these methods have limitations in addressing the complexity of biological systems [7]. In contrast, deep learning models can accommodate the non-linearity and interaction among loci in complex biological traits and could potentially mitigate the limitation of the classic genome-wide association method [8].

Recently, a linear regression based GWAS on SARS-CoV-2 fitness identified 2904 amino acid changes associated with fitness changes [9]. A few fitness associated amino acid changes were concentrated in the Spike gene region, some of which were further validated by experiments. Some fitness associated mutations were also found concentrated in N, M, ORF1a and ORF1b genes in SARS-CoV-2. Another regression based GWAS identified 44 non-synonymous changes in SARS-CoV-2 associated with virion copy numbers – a proxy of viral fitness [10]. Although helpful for our understanding of the SARS-CoV-2 evolution, both regression-based GWAS did not consider the potential epistatic interactions among loci associated with fitness changes.

A common approach in estimating the pandemic impact of SARS-COV-2 variants is the WHO designation of the variants. The WHO designates a SARS-CoV-2 variant as a VOC if it has genetic changes affecting transmissibility, virulence, or treatment efficacy and shows a growth advantage[11]. A VOC, through a WHO risk assessment, also demonstrates significant clinical severity, major epidemiological impact on health systems, or a marked reduction in vaccine effectiveness[11].

To address the complexity of linking genotypes to phenotypes for SARS-CoV-2, we explored machine learning approaches, including deep learning, as alternative for genome-wide association studies. We estimated SHAP values for genomic features to improve model interpretability. SHAP assigns a value to each feature, representing its impact on the model's output. We report that convolution neural network model provides biologically relevant and accurate prediction of mutations associated with SARS-CoV-2 VOC labels in comparison with other three machine learning models. We compare the associated mutations identified by explainable CNN model with those found by traditional GWAS. Our findings showed that the explainable CNN approach can identify biologically relevant mutations by capturing non-linear interactions, offering a more relevant understanding of DNA sequencing of SARS-CoV-2.

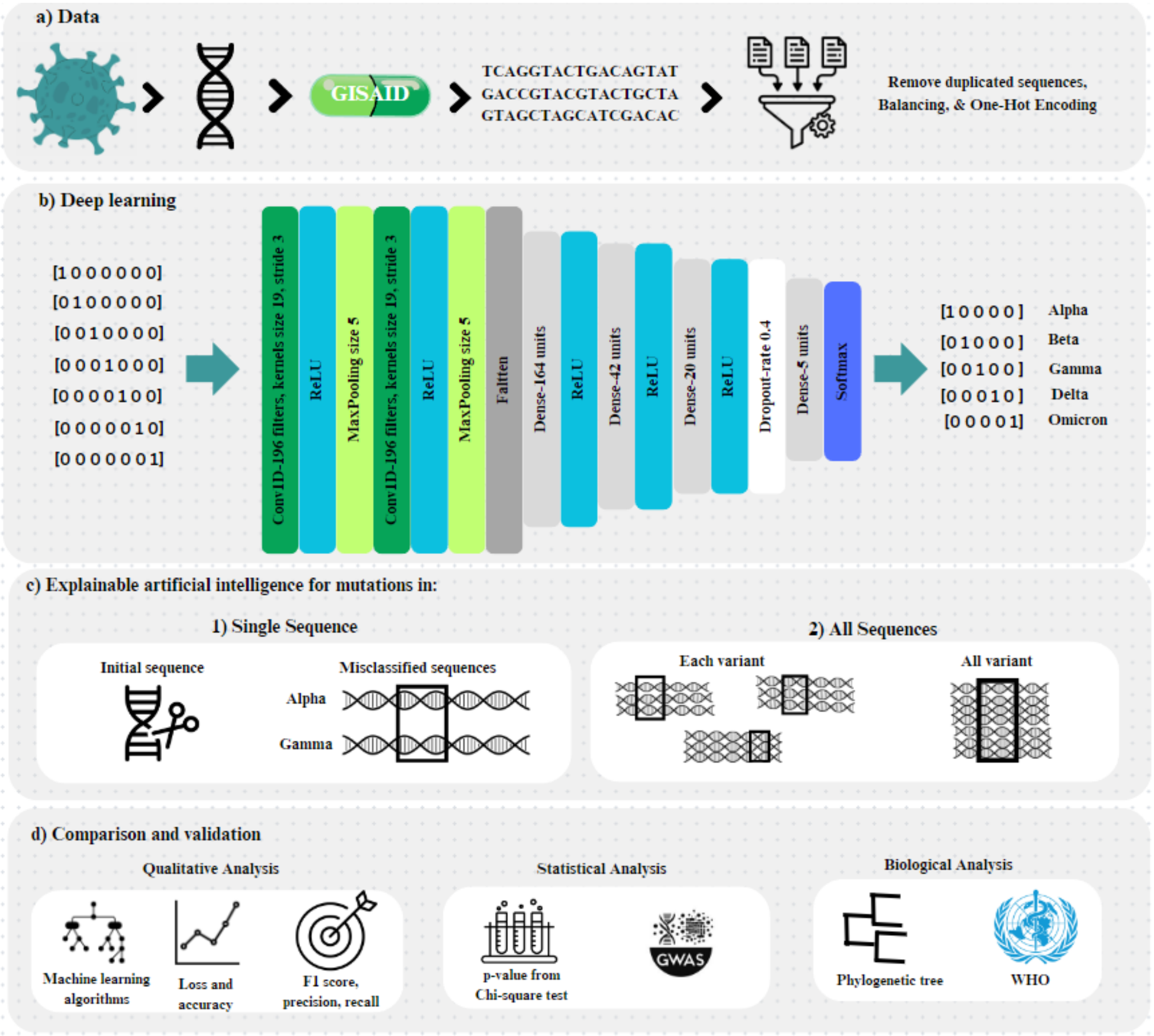

***Figure 1***. *An overview of this study*

# Results

## CNN is an effective machine learning model for SARS-CoV-2 Classification

We evaluated four machine learning approaches for predicting WHO labels from input sequences: three tree-based models—Random Forest[12], XGBoost[13], and CatBoost[14]and a 2-layer Conv1D model (Fig. 2). SHAP value analysis indicates that the 2-layer Conv1D model is better at identifying biologically relevant regions compared to the tree-based models. Consequently, we focus subsequent studies on the 2-layer Conv1D model.

A summary of this study in presented in Fig.1. The input SARS-CoV-2 genomes was downloaded from the Global Initiative on Sharing Avian Influenza Data (GISAID)[15]. This study focus on genomes from VOCs: Alpha, Beta, Delta, Gamma, and Omicron[11]. After cleaning and balancing of data, we obtained a total of 200,000 sequences of SARS-CoV-2. One-hot encoding was applied to all input sequences. We partitioned the dataset into three subsets: 128,000 sequences for training, 32,000 for validation to fine-tune model parameters, and 40,000 for testing to assess the model's predictive accuracy. We implemented 2-layer Conv1D classification model due to its ability to effectively extract sequence-specific features just along one dimension.

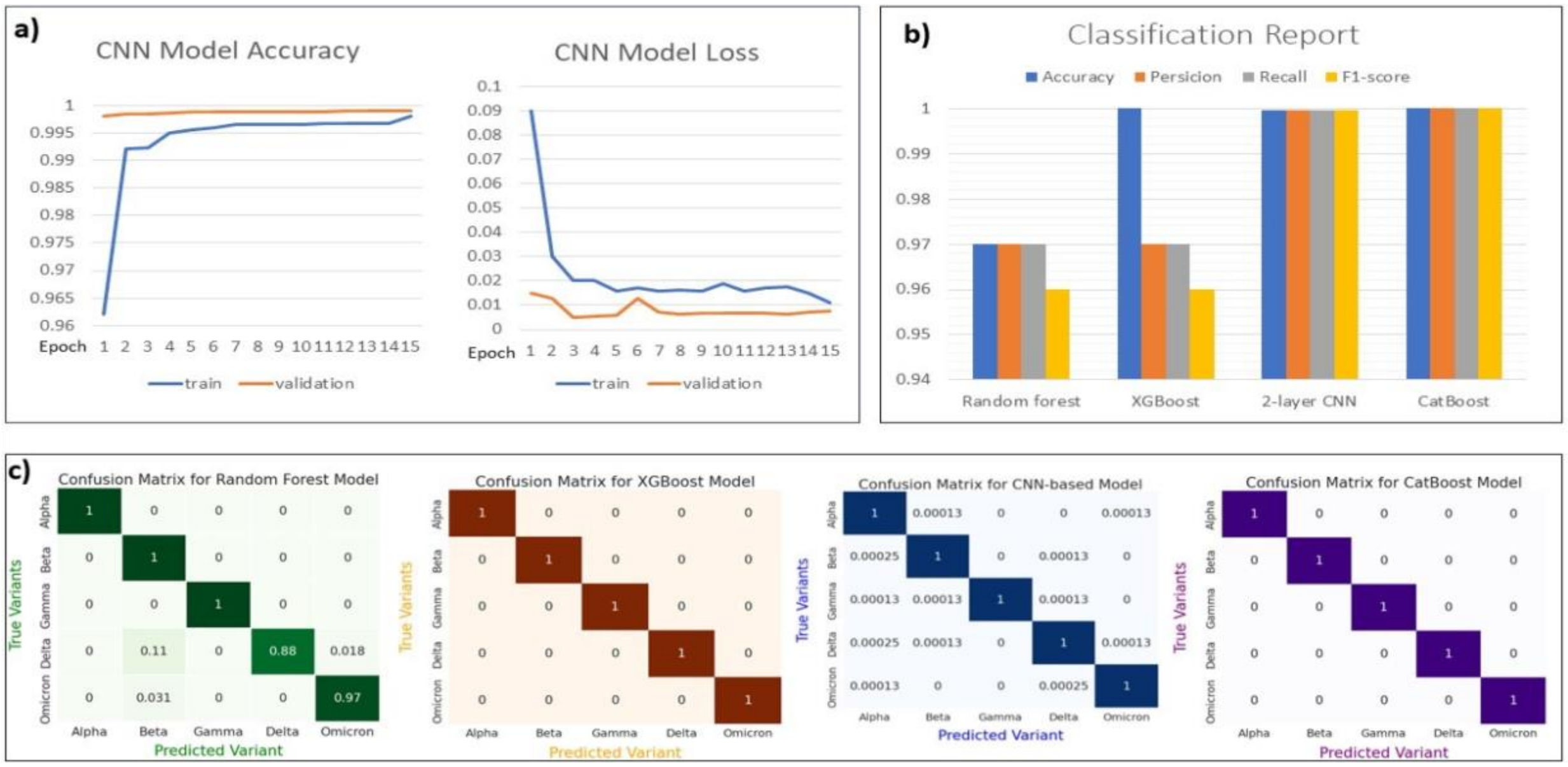

***Figure 2**. Performance overview of four machine learning approaches. Based on accuracy, precision, recall, F1 Score, and Confusion matrix, CNN-based model and CatBoost can classify the genomic sequence better in comparison to Random Forest and XGBoost.*

All four models underwent optimization procedures. Both the 2-layer Conv1D model and CatBoost exhibited better performance in comparison to Random Forest and XGBoost (Fig.2b). In the 2-layer Conv1D model, the loss on the training data is 0.0111 and the validation loss is 0.0024, and the accuracy on test data set is 99.96%.

## The Importance of Specific Mutations from SHAP Value Analysis

To determine which genomic positions are important for model predictions, we performed SHAP value analysis on individual input sequences. SHAP identifies important features as those with high absolute Shapley values. The base value of SHAP estimation for each label were derived using a subset of five sequences randomly selected from the training data. Subsequently, we selected a reference genome each variant: Alpha (EPI_ISL_60144[16]), Beta (EPI_ISL_712073[16]), Delta (EPI_ISL_3148365[16]), Omicron (EPI_ISL_6640916[17]), and Gamma (EPI_ISL_2095177) to calculate their SHAP values for each position. We focused on mutations that leading to amino acid changes, especially in the Spike gene SARS-CoV-2[18], and examine whether results of Shapely value analyses align with the mutations used to define each variant by WHO[19] (Table 1). [18]

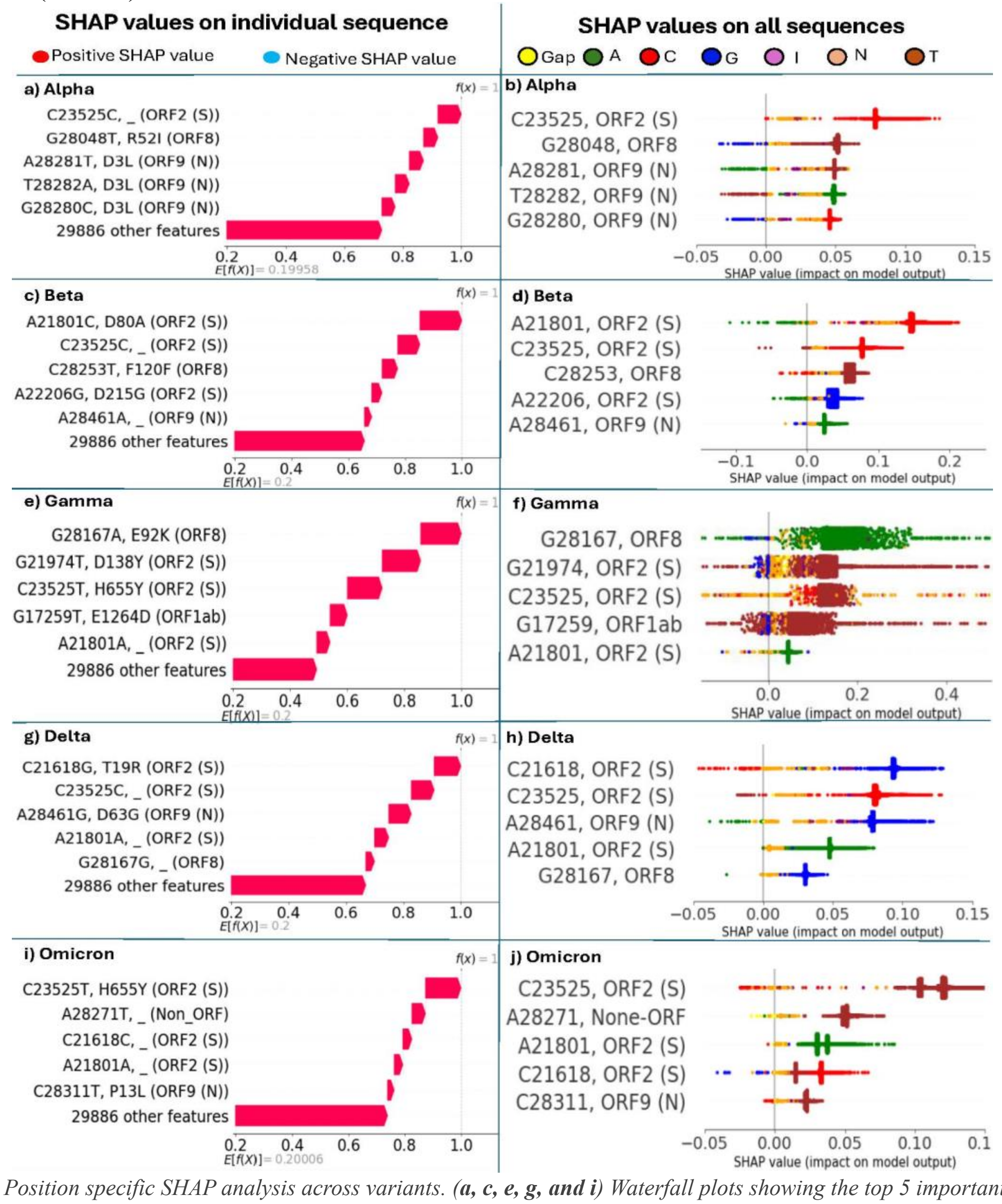

***Figure 3***. *Position specific SHAP analysis across variants. (**a, c, e, g, and i**) Waterfall plots showing the top 5 important positions in the initial or random genetic sequences for each variant. Each step in the plot represents the impact of an individual position on the model's prediction, where the direction and size indicate the influence's magnitude. These plots begin with the model's base value (E(f(x))) and sequentially add the impact of each position's SHAP value, illustrating the cumulative effect on the prediction. Note: All steps in these samples are positive, increasing the prediction value. (**b, d, f, h, and j**): Beeswarm plots representing the aggregated SHAP values used to assess the global significance of each position across all sequences within each variant. This technique calculates the mean absolute SHAP values for each position, ordering features on the y-axis based on their impact. Points to the right indicate a positive influence on the model's output, while those to the left indicate a negative impact. The color of each point indicates the nucleotide type.*

*Table.1 Top 5 Important Genomic Positions in Each SARS-CoV-2 Variant for CNN model as determined by SHAP. Genomic positions include corresponding nucleotide and/or amino acid changes, gene*, and their representations.

| Variant | Nucleotide Position | Condon | Amino acid changes | ORF | Mutation in Reference Sequence | Relevance to known markers of Variants |
|---|---|---|---|---|---|---|
| Alpha (Fig.3.a, Fig.3.b) | C23525C | - | - | ORF2 (S) | No[16] | Gamma (H655Y *)[20], Omicron (H655Y**)[21] |
| | G28048T | AGA>ATA | R52I | ORF8 | Yes[16] | - |
| | A28281T | GAT> CTA | D3L | ORF9 (N) | Yes[16] | - |
| | T28282A | GAT> CTA | D3L | ORF9 (N) | Yes[16] | - |
| | G28280C | GAT> CTA | D3L | ORF9 | Yes[16] | - |
| Beta (Fig.3.c, Fig.3.d) | A21801C | GAT>GCT | D80A* | ORF2 (S) | Yes[16] | - |
| | C23525C | - | - | ORF2 (S) | No[16] | Gamma (H655Y *)[22], Omicron (H655Y**)[21] |
| | C28253T | TTC>TTT | F120F | ORF8 | No[16] | Sub lineage of Beta[23], Sub lineage of Gamma (F120F)[24] |
| | A22206G | GAT>GGT | D215G* | ORF2 (S) | Yes[16] | - |
| | A28461A | - | - | ORF9 (N) | No[16] | Delta (D63G)[16] |
| Gamma (Fig.3.e, Fig.3.f) | G28167A | GAA>AAA | E92K | ORF8 | Yes[22] | - |
| | G21974T | GAT>TAT | D138Y* | ORF2 (S) | Yes[22] | - |
| | C23525T | CAT>TAT | H655Y * | ORF2 (S) | Yes[22] | Omicron (H655Y**)[21] |
| | G17259T | GAG>GAT | E1264D | ORF1ab | Yes[22] | - |
| | A21801A | - | - | ORF2 (S) | No[22] | Beta (D80A*)[16] |
| Delta (Fig.3.g, Fig.3.h) | C21618G | ACA>AGA | T19R* | ORF2 (S) | Yes[16] | - |
| | C23525C | - | - | ORF2 (S) | No[16] | Gamma (H655Y *)[22], Omicron (H655Y**)[21] |
| | A28461G | GAC>GGC | D63G | ORF9 (N) | Yes[16] | - |
| | A21801A | - | - | ORF2 (S) | No[16] | Beta (D80A*)[16] |
| | G28167G | - | - | ORF9 (N) | No[16] | Gamma (E92K) |
| Omicron (Fig3.i, Fig.3.j) | C23525T | CAT>TAT | H655Y** | ORF2 (S) | Yes[21] | Gamma (H655Y *)[22] |
| | A28271T | - | - | between ORF8 and ORF9 (N) | Yes[25] | - |
| | C21618C | - | - | ORF2 (S) | No[21] | Delta (T19R*)[16] |
| | A21801A | - | - | ORF2 (S) | No[21] | Beta (D80A*)[16] |
| | C28311T | CCC>CTC | P13L* [26] | ORF9 (N) | Yes[21] | - |

*: Mutation in spike protein that has been identified by WHO[26,27]
**: Mutation in Omicron that has been identified by Coronavirus Antiviral & Resistance Database in Stanford University[28]
ORF: open reading frame, del: nucleotide deletion
-: No mutation

In the analysis of SHAP values derived from our model, as detailed in Table.1, there is a general agreement between the top-ranked SHAP values, and the mutations found in the reference sequences of each VOCs.

Notably, position 23525 consistently exhibits high SHAP values across all VOCs, and position 21801 across four VOCs. These two positions correspond to the amino acid changes H655Y in Gamma[22] and Omicron[21] and D80A in Beta[27], both located within the spike gene. The C23525T H655Y mutation is associated as increased fitness of the SARS-CoV-2 variant [29]. The A21801C D80A has a sever effect on immune system[30]. Both H655Y and D80A are acknowledged by the WHO as contributing to the emergence of the Gamma and Beta variant[27] respectively. The presence of nucleotide "T" at position 23252 in the Gamma (Fig.3f) and Omicron (Fig.3j) sequences and "C" at position 21801 in Beta (Fig.3d) sequences demonstrate a positive influence on predicting the correct labels. Interestingly, in the reference sequence where position 23252 contains nucleotide "C" and position 21801 contain "A", in the absence of any mutations at this site, the deep learning model learn to use this feature to predict the right VOCs (Fig.3b, Fig.3d, Fig.3f, Fig.3h, Fig.3j).

For the Alpha variant (Fig. 3a and 3b), the second highest SHAP value corresponds to the G28048T (R52I: ORF8) mutation, which has been linked with possible interaction with host immune system [31] (Fig.3a). The presence of 'T' at position 28048 strongly supports the identification of Alpha label based on Shapely values (Fig. 3b). The mutations G28280C, A28281T, and T28282A, leading to the D3L amino acid change[32] that is prevalent in 98.4% of Alpha variant samples[33]. The "C" in position 28280, "T" in position 28281, "A" in position 28282 are strong positive features for classifying Alpha (Fig.3b).

For the Beta variant (Fig. 3c and 3d), two significant mutations identified by WHO in the spike protein of the Beta variant influence the model's predictions: A21801C (D80A: ORF2 (S)), located in the N-terminal domain (NTD) of the SARS-CoV-2 spike protein[34], and A22206G (D215G: ORF2 (S)). These two mutations are indicated by the highest positive SHAP values in a reference sequence of Beta (Fig.3c). These two mutations have a significant role in weakening the immune system[30]. The presence of "G" at position 22206 has a positive effect on the Beta label. Conversely, the presence of "A" at this position- when there is no mutation in this position- guides the model to classify the sequence as belonging to another VOC besides Beta (Fig.3d). The third important mutation for Beta is C28253T (F120F: ORF8) [23] (Fig.3c)[23]. When there is "T" in position 28253, it leads models toward Beta, whereas a "C" would classify the sequence as another variant (Fig.3d). Interestingly, the fifth SHAP value is for position 28461(Fig.3c), which corresponds to a specific mutation to Delta. The model learned to use this wildtype 'A' residue as a feature for Beta (Fig.3d).

For the Gamma variant (Fig. 3e and 3f), the mutation G28167A (E92K: ORF8), the unique mutation occurring in ORF8 within the lineage of Gamma [35], exhibits the highest SHAP value in the model. The mutations G21974T (D138Y: ORF2 (S)) and C23525T (H655Y: ORF2 (S)) are known markers by WHO for Gamma [27]. The presence of "T" at these two positions has mostly positive effects on model's ability to identify Gamma. Additionally, the mutation G17259T (E1264D: ORF1ab) ranks fourth in SHAP values, and "T" at this position guides the model toward Gamma (Fig.3f).

For the Delta variant (Fig. 3g and 3h), the most influential mutation is C21618G (T19R: ORF2 (S)), a key mutation in the spike protein recognized by the WHO as characteristic of Delta[27] and likely involved in the immune responses to the virus[36]. Another mutation is A28461G (D63G: ORF9 (N)), which is associated with increased risk of intensive care unit (ICU) admission or death [37]for patients [37]. The "G"s in positions 21618 and 28461 increase the chance of input sequence to be classified as Delta (Fig. 3h). The position G28167G correspond to the G28167A mutation for Gamma, and it shows that the model learns to use the wildtype residue for an important feature for proper classifications.

For the analyzed Omicron example (Fig. 3i and 3j), the SHAP values highlight the mutations C23525T (H655Y: ORF2 (S)), A28271T (non-ORF), and C28311T (P13L: ORF9), which are known previously for the Omicron variant[28].

For comparison, we provided SHAP value plots for CatBoost approach in the supplementary document (Fig. S1, Table. S1). In addition, we further analyze the causes of CNN misclassifications for selected examples provided in the supplementary document (Fig. S2–Fig. S6).

## Impact of Mutations on Differentiating Variants

We present heatmaps to illustrate that when a genomic position with mutation is associated with one variant label, the same position with different nucleotide residues is often associated with different variants (Fig.4A). For example, when the input is Alpha, positions 28280, 28281, and 28282 positively influence the model's decision to identify the input as Alpha. At the same time, these positions often negatively impact the model's prediction for other variants. The same trends can be seen in many other positions within each variant.

We illustrated some of them with the Wuhan-Hu-1 reference and mutated nucleotides with magnitude of SHAP value in Fig.4B, which correspond to highlighted by yellow box in Fig.4A. Each panel provides a segment of the reference sequence where these mutations occur, alongside the associated SHAP values that are represented by the direction and magnitude of each letter. It follows the pattern of positive SHAP values supporting correct classification and negative values indicating other labels. In all panels of Fig. 4B, the second row represents feature that has important effect for the true variant label, and the bottom row with mirror-flipped letters indicates their effect on other variant labels.

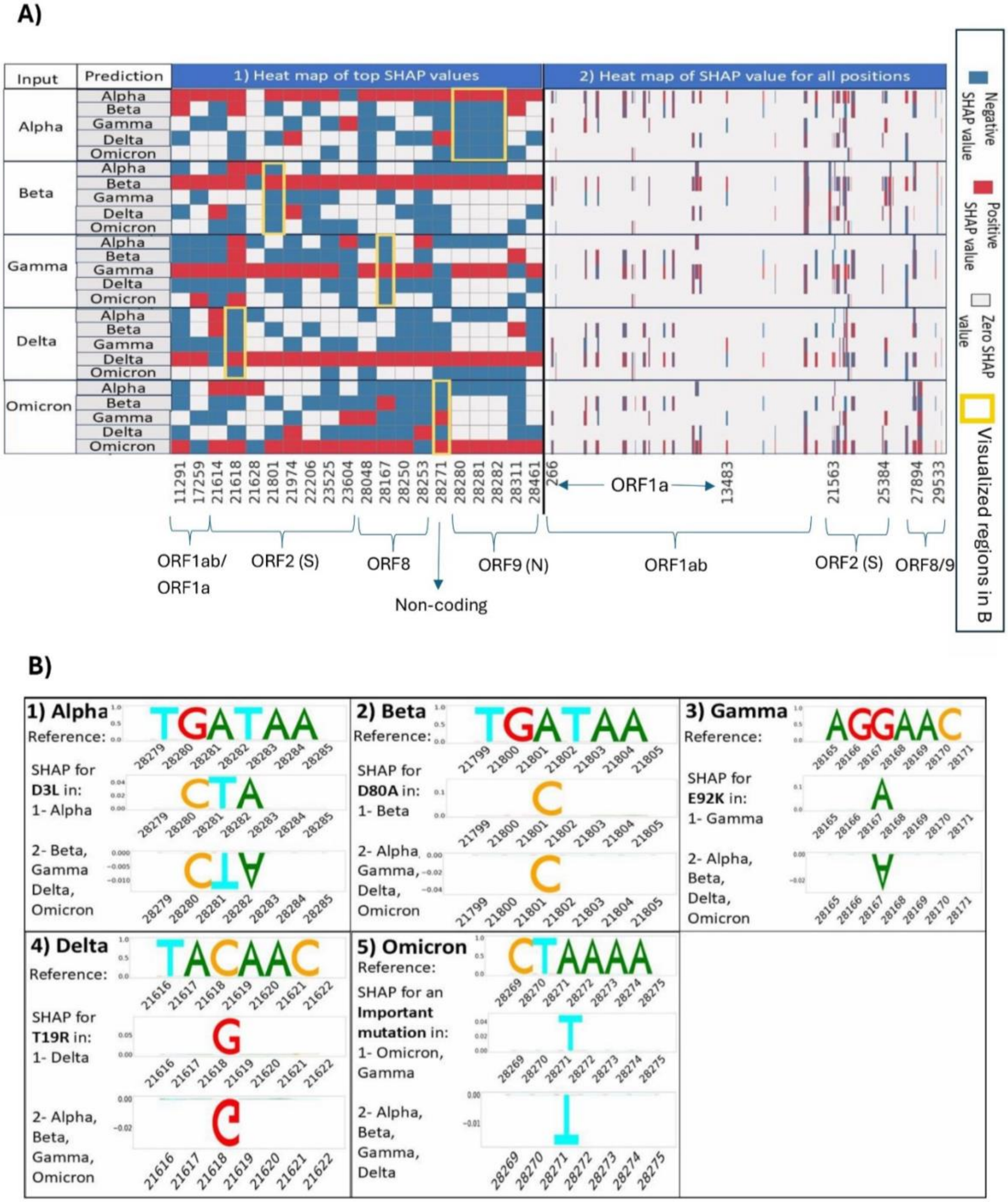

*Figure 4. Visualization of SHAP values in example sequences. A) Regions highlighted in red indicate positive SHAP values, increasing the likelihood of accurate classification to the variant's true class. Conversely, blue regions indicate negative SHAP values, reducing the likelihood of misclassification to other classes. A.1 displays the heatmap of SHAP value signs for regions with the highest absolute SHAP values within each variant, while A.2 shows SHAP value signs across all genomic positions in the sequence for each variant. Yellow boxes indicate key mutations within each variant, and their SHAP values are illustrated in B. B) The impact of each mutation is visualized, showing its positive effect on correctly classifying the sequence to its own class and its negative effect on misclassifying the sequence into other variants. Within each block, the first sectional sequence is the Wuhan-Hu-1 reference sequence, the second sectional sequence shows the sign (positive) and magnitude of mutations from input sequence when model classifies it correctly, and the third section illustrates the sign (negative) and magnitude of those mutations if the model assigns the input sequence to other variants incorrectly.*

## Comparison between xAI and GWAS on identify mutations associated with VOCs

Genome-wide association studies (GWAS) are a standard method for identifying mutations linked to phenotypic changes [38]. Comparing the xAI approach with GWAS can highlight how machine learning techniques complement traditional biostatistical methods. We implemented a GWAS using a chi-square-based association analysis and then applied a -log10 transformation to the p-values. For comparison, [38]we calculated aggregated SHAP values (Eq.6), which measure the impact of each genomic position on our model's predictions across all VOCs.

We first compared the SHAP results with -log(p-values) from the GWAS using scatter plots (Fig. 5a and 5b). Notably, the Shapley value analysis clearly highlight some mutations in the spike region (Fig. 5a), in contrast to the much noisy picture generated by GWAS (Fig. 5b). [39]To quantitatively assess the contribution of each genomic regions, we analyzed the summation of -log(p-values) and aggregated SHAP values across all genes individually (Fig. 5c). In GWAS, the ORF1ab region contributed 38% of the top-ranked -log10 p-values. In the Shapley value analysis, the spike region (ORF2 (S)) accounts for 50% of the highest aggregated SHAP values. Given the prominent role of spike region in the definition of VOCs, the results here suggest that the SHAP results are more biologically relevant to VOCs than GWAS.

We then zoom-in on regions with high values to further examine which genes are most impactful (Fig. 5d). We focused on the 1,394 positions (considering ORF1ab and ORF1a individually) with a normalized -log(p-value) of closer to 1 (Fig.5d). We found that ORF1AB, ORF1A, and ORF2(S) contribute 24%, 23%, and 29%, respectively, to the highest normalized -log(p-value), while the spike region (ORF2(S)) is particularly dominant in SHAP contributions, accounting for 55% of the highest values.

We next present Venn diagrams to illustrate complementary performance between SHAP and GWAS methods by focusing on the top positions from both methods (Fig. 5e, 5f, and 5g). To address potential sensitivity of threshold choice, we picked top 1073, 1500, and 2000 positions from both methods. First, we consistently observe overlapping top features selected by both methods, 23.8% ~ 32.4%. Furthermore, majority of these overlapping features, 53.3~67.6%, are from the Spike gene. Second, we also observe large fraction of distinct top features selected by either GWAS or SHAP, suggesting that SHAP offers a different perspective on the link between genotype and phenotype to GWAS. In the distinct groups of features, the influence from the Spike genes is not as pronounced as that in the intersected group.

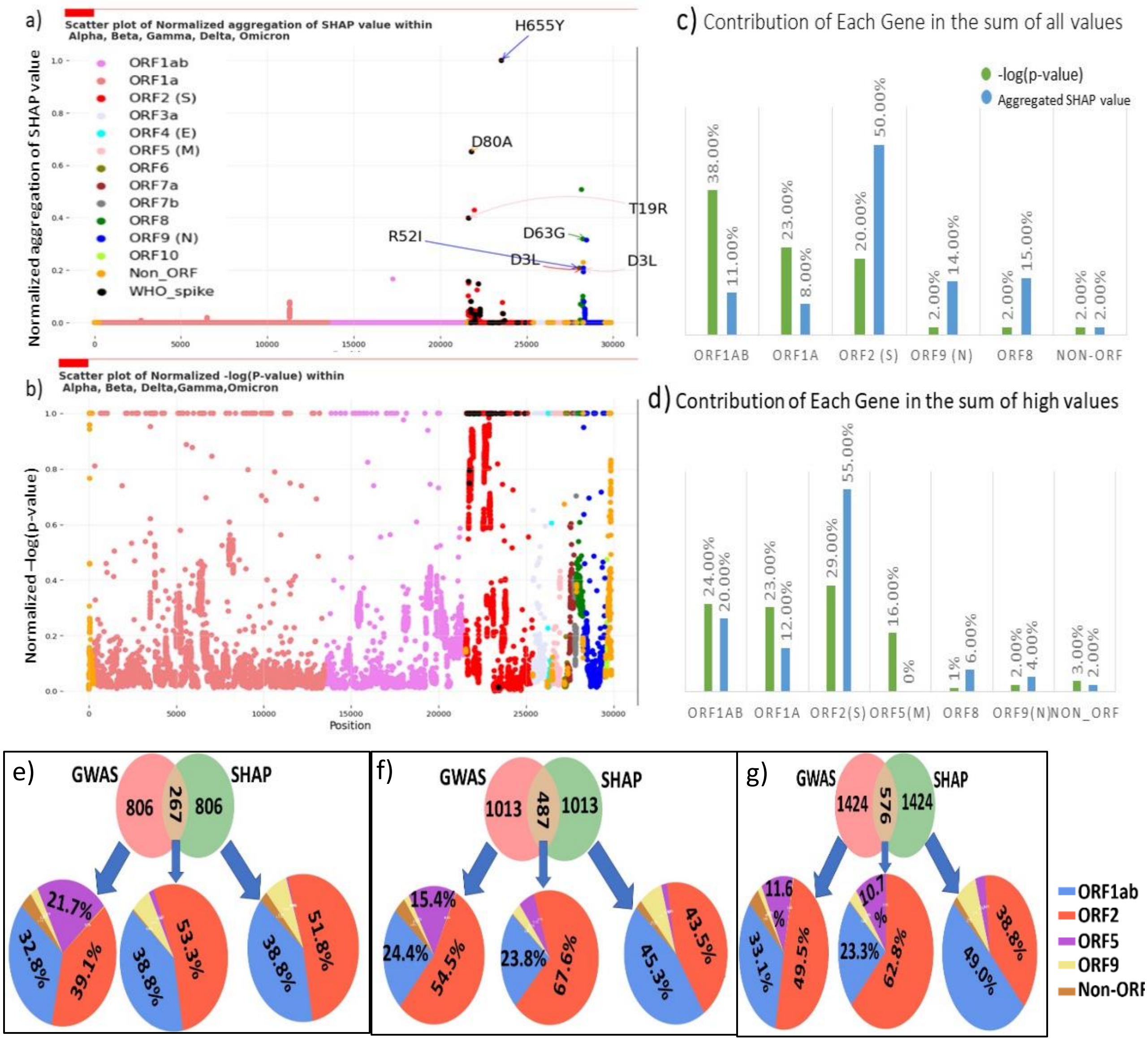

*Figure 5. Comparison of Gene Importance and Positional Contributions between SHAP and GWAS. Comparison of values in high normalized -log(p-value) (a) and normalized aggregated SHAP values (b) across all genes and variants. (c) Highlights the importance of each gene for GWAS and 2-layers CNN models, while (d) shows which genes with high values have greater importance in these analyses. in e, f, and g the number of positions with high values in normalized -log(p-values) and normalized SHAP values are compared. (e), (f), (g) Pie charts comparing the distribution of gene contributions among 1073, 1500, and 2000 top positions for GWAS (-log(p-value)) and SHAP values. These charts show the proportion of high-value positions attributable, in GWAS, SHAP, and intersection of them, to specific genes*

# Discussion

In this study, we combined a convolutional neural network with post hoc explanation techniques[39] to identify biologically significant positions and compare them with those from GWAS. We found that the explainable convolutional network approach can infer known mutations, especially in the spike gene, more effectively than the classic GWAS. Furthermore, the explainable convolution neural network consistently identified mutations in ORF8 and ORF9 and other mutations that often eluded the classic GWAS. These findings suggest that the explainable convolutional neural network is an alternative approach to GWAS to identify potential genetic changes associated with phenotypic changes.

Our results indicate that genomic positions linked to variant-specific mutations have a significant impact on the convolutional neural network model's performance. For example, top-ranking markers include R52I and D3L for Alpha; D80A, F120F, and D215G for Beta; T19R and D63G for Delta; H655Y for Omicron; and E92K, D138Y, P681R, H655Y, and E1264D for Gamma sequence [19,40]. Notably, the wildtype reference in these positions (absence of these mutations) are also important features identified by SHAP analysis, such as 23525 and 21801 for Alpha; 23525 and 28461 for Beta; 21801 for Gamma; 23525, 21801, and 28167 for Delta; and 21618, 21801, and 28461 for Omicron. Interestingly, SHAP analysis did not prioritize the well-known mutation A23403G (D614G: ORF2 (S)) because it is present across all studied variants and hence is not an effective feature for classification.

We also examined whether SHAP analysis is related to the classic phylogenetic approach [41]. We performed a phylogenetic tree analysis using multiple genetic sequences of each VOC. We then compared the phylogeny with the hierarchical clustering tree based on the correlation distance matrix derived from the SHAP value (Fig. S9). The SHAP-based tree shares partial similarity with the phylogenetic tree, which further supports the biological relevance of the proposed explainable convolutional network models.

The current work uses VOC label as a proxy of the viral transmission phenotype. More direct measurements of the SARS-CoV-2 transmission or pathogenic phenotypes include the viral load [42], fitness estimation [43], and patient mortality risks [44], which would require much more resources and may be addressed in future studies. As with most supervised machine learning approaches, the explainable convolutional neural networks approach requires correctly annotated labels for the input sequences.

Overall, this study provides strong evidence that convolutional neural networks can effectively capture nucleotide changes that are associated with phenotypic changes and can outperform classic GWAS in some aspects, likely because convolutional neural networks can accommodate non-linear interactions among genetic loci. Machine learning has great potential in public health, such as the prevention of infectious diseases [3,45,46,] forecasting [4], diagnosis [47,48], policy development [49] , and hospital resource allocation and preparation [50]. We hope this work will lead to further research into leveraging machine learning to address challenges in public health.

# Materials and Methods

## Data and its Preprocess

All data used in this study was downloaded on 06/16/2022 from GISAID[15]. The data contains 11,378,408 genetic sequences of the SARS-CoV-2 virus, each composed of 29,891 nucleotides. These variants have been assigned specific names as recommended by WHO. For this study, we consider Alpha, Beta, Delta, Gamma, and Omicron as VOCs. We down-sampled the data to address the imbalance issue, which resulted in a dataset with 200,000 genetic sequences of SARS-CoV-2 in VOCs. We used one-hot encoding to represent each nucleotide sequence as a 7-dimensional matrix. Each dimension corresponds to a specific base (A, C, G, T), the gap, ambiguous nucleotides (N), and other nucleotides, denoted as "i". A "1" is assigned to the dimension corresponding to the base in the sequence, while the rest are "0." These 7-dimensional vectors are concatenated along the sequence, resulting in a 200,00× 29,891 × 7 matrix. 160,000 sequences were allocated for training and validation and 40,000 testing.

In tree-based models, we sampled 300 genetic sequences from each variant, resulting in 1500 sequences. Out of these, 1200 were used for training and validation, and the remaining 300 were set aside for testing. We flattened the one-hot encoding matrices into two-dimensional matrices. This resulted in a data dimension of 1200 rows by 209,237 columns for training and 300 rows by 209,237 columns for testing. For target labeling, we used label encoding.

## Training of the 2-layer CNN model

We trained the CNN-based model for 15 epochs with the Adam optimizer with a learning rate of 0.0014924, and an effective batch size of 64. Hyperparameters of the model, such as the number of filters, kernel sizes, strides, activation functions, and learning rate were fine-tuned through a grid search procedure. Categorical cross-entropy served as the loss function during training. The training computations were executed on an NVIDIA GeForce RTX 3090.

## Shapley Additive Explanation (SHAP)

We used Deep SHAP, an advanced version of SHAP, due to its superior capability in interpreting predictions made by deep neural networks[6]. Each sequence in our study comprised 29,891 features, each represented as a 7-dimensional array due to one-hot encoding. This encoding transforms each nucleotide (feature) into a 7-dimensional array, resulting in a disparity between the SHAP values and the multi-dimensional nature of the features. To address this, we aggregated the SHAP values across all seven dimensions for each nucleotide for each RNA sequence.

We also employed two properties of SHAP to analyze genetic sequences locally:

1. Property 1(Additive Feature Attribution in SHAP)
   The corresponding formula is as follows[6]:

   $$g(x') = \varphi_0 + \sum_{i=1}^{M} \varphi_i x'_i \qquad (3)$$

   In Eq.3, $g(x')$ represents an explanation model, serving as an interpretable approximation of our original predictive model. The variable $x'$ represents a transformed or simplified version of the input, essentially mapping the original input to a format more suitable for the explanation model $g(x')$. $M$ is the number of features of mapped input. This transformation is expressed as:

   $$x = h_x(x') \qquad (4)$$

   Where $h_x$ is the function that maps $x'$ to $x$. It's important to note that $x'$ serves as a binary representation of the features, indicating the presence or absence of each feature. The term $\varphi_0$ denotes the base value, reflecting the model's output with all simplified inputs toggled off, or based on Eq.4 it is:

   $$\varphi_0 = f\big(h_x(0)\big) \qquad (5)$$

2. Property 2 (Local Accuracy):
   This property leverages from additive feature attribution. The main idea of Eq.3 is to approximate the function $g(x')$ to $f(h_x(x'))$, where $f(x)$ is our original model. $f(x)$ is designed to attribute an effect, represented by $\varphi_i$ to each feature.

   Eq.3 symbolizes the average effect of all features, ensuring that the summation of all SHAP values and the base value equates to the output of the original model for each data[6]. In essence, the concept of local accuracy in the context of SHAP values implies that these values offer a precise and context-specific interpretation of the model's prediction for any given input. Eq.3 enabled us to find sequences classified correctly based on their SHAP values and subsequently analyze them in-depth.

In our study, we applied these two properties to assess genetic sequences for SHAP value analysis. According to the properties, when we add the base value with the sum of SHAP values for all features, if simple models ($g(x')$) can accurately classify the simplified input ($x$) like original model $f(x)$, it suggests that the simple model performs well and like original model on the sample. This, in turn, validates the SHAP values generated, demonstrating that SHAP values completely correspond to the importance of each feature in the original model.

For global feature importance, we employed the practice of averaging the absolute Shapley values associated with each feature across the dataset[37].

$$I_j = \frac{1}{n} \sum_{i=1}^{n} |\varphi_j^{(i)}| \quad (6)$$

### GWAS on mutations associated with WHO labels

Genome-wide association study (GWAS) is a method to identify genetic variations that correlate with a specific trait of a disease. Here, we used a chi-square test to examine if certain mutations in genes were significantly associate with VOCs. To facilitate comparison with the SHAP value, we first transformed the p-values with a negative logarithmic function $-\log(P_{value})$. Both the aggregated SHAP values and $-\log(P_{value})$ were then normalized using min-max scaling to standardize the values into a range from 0 to 1.

### Data and Code Availability

GISAID data can be accessed at https://gisaid.org/. Codes are available at https://github.com/QinLab/SARS-DeepLearning2024.

## Acknowledgment

HQ and LQ thank the NSF award 2200138. PH, HQ and MX acknowledge the support of NSF award 2234910. HQ thanks the internal support at the Old Dominion University. We acknowledge the support of high-performance computing facility at University of Tennessee at Chattanooga and Old Dominion University. LUM and JG are partially supported by NSF REU 2149956.

## Author Contribution

HQ conceived and designed the project. HQ, LQ, and MX secured funding support. PH conducted the research with assistance from RA, LUM and JG. PH drafted the manuscript. HQ, LQ and MX revised the manuscript.

## Competing interests

The authors declare no competing interests in this project.

Supplementary Materials

# Explainable convolutional neural network model provides an alternative genome-wide association perspective on mutations in SARS-CoV-2

Parisa Hatami, Richard Annan, Luis Urias Miranda, Jane Gorman, Mengjun Xie, Letu Qingge, Hong Qin

## Table of Contents

## 1. SHAP value  Catboost

In this study, we trained multiple models, including XGBoost[1], Random Forest[2], CatBoost[3], and a 2-layer CNN model, for classification tasks. Among these models, CatBoost and the 2-layer CNN demonstrated higher accuracy. To understand the decision-making processes of these models, we analyzed the initial sequences of Alpha, Beta, Delta, Omicron, and a random sequence from Gamma. By calculating the SHAP values for these sequences, we observed that the 2-layer CNN model relies more heavily on significant mutations (Figure 3 and Table S1) compared to CatBoost (Figure S1 and Table S1).

Another limitation of CatBoost is its inability to directly train on 3-dimensional data. Consequently, we had to reshape our data into two dimensions. This transformation resulted in a total of 209,237 features instead of the original 29,891, as each feature expanded to seven dimensions due to one-hot encoding. Each position in the dataset, therefore, has seven SHAP values, indicating which nucleotide is more influential at that position. As illustrated in Supplementary Table S1, some positions have two or more nucleotides with the highest SHAP values, highlighting their relative significance.

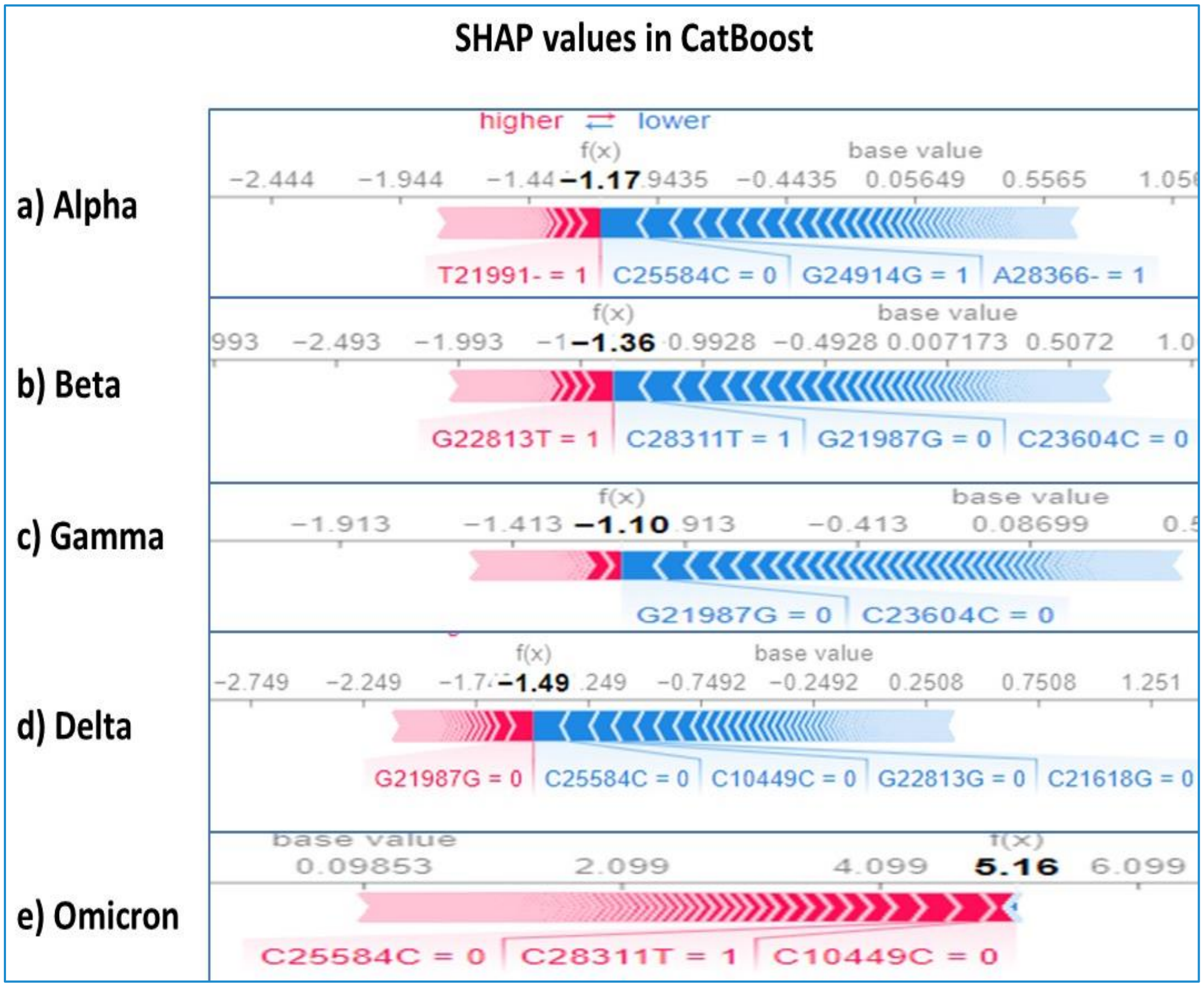

Figure S1. SHAP Values in CatBoost: Like CNN, CatBoost exhibits high accuracy in classifying RNA sequences of SARS-CoV-2. Analysis of CatBoost reveals that its classification is not primarily based on significant mutations within variants. Supplementary Table S1 outlines the positions prioritized by CatBoost in the classification process. Sequences used in this table include Alpha (EPI_ISL_60144[4]), Beta (EPI_ISL_712073[4]), Delta (EPI_ISL_3148365[4]), Omicron (EPI_ISL_6640916[5]), and a random sequence from Gamma (EPI_ISL_2095177).

Supplementary Table S1. Although CatBoost has high accuracy in the classification of genomic sequences, it is not able to recognize the important mutations for each variant

| **Variant instances GISAID ID** | Nucleotide Position | Amino acid changes | ORF | Mutation in Initial Sequence | Presence in Initial Sequence of Other Variants or Sub lineage of the Variant |
|---|---|---|---|---|---|
| Alpha (Fig.S.1.a)<br>Id: EPI_ISL_60144 | C25584C | - | ORF3a | No[4] | - |
| | G24914G | - | ORF2 (S) | No[4] | Alpha (G24914C: D1118H)[4] |
| | T21991- | Y144del | ORF2 (S) | Yes[4] | - |
| | A28366- | - | ORF9 | No[4] | - |
| | C10449C | - | ORF1ab | No[4] | - |
| Beta (Fig. S1.b)<br>Id: EPI_ISL_712073 | C28311T | - | ORF9 | No[4] | - |
| | G21987G | - | ORF2 (S) | No[4] | Delta (G21987A: G142D)[4] |
| | G22813T | K417N | ORF2 (S) | Yes[4] | - |
| | C23604C | - | ORF2 (S) | No[4] | Delta (C23604G: P681R)[4] |
| | T22282T | - | ORF2 (S) | No[4] | Beta (Δ 22281–22289: LAL 242–244 del)[4] |
| Gamma (Fig. S1.c)<br>Id: EPI_ISL_2095177 | G21987G | - | ORF2 (S) | No[6] | Delta (G21987A: G142D)[4] |
| | C23604C | - | ORF2 (S) | No[6] | Delta (C23604G: P681R)[4] |
| | C23604A | - | ORF2 (S) | No[6] | Delta (C23604G: P681R)[4] |
| | G22813T | K417N | ORF2 (S) | No[6] | Beta[4] |
| | G22813G | - | ORF2 (S) | No[6] | Beta (G22813T: K417N)[4] |
| Delta (Fig. S1.d)<br>Id: EPI_ISL_3148365 | C25584C | - | ORF3a | No[4] | - |
| | C10449C | - | ORF1b | No[4] | - |
| | G21987G | - | ORF2 (S) | No[4] | Delta (G21987A: G142D) [4] |
| | G22813G | - | ORF2 (S) | No[4] | Beta (G22813T: K417N) [4] |
| | C21618G | T19R | ORF2 (S) | Yes[4] | - |
| Omicron (Fig. S1.e)<br>Id: EPI_ISL_6640916 | C10449C | - | ORF1ab | No[5] | Unique common mutation in Omicron (C10449A)[5] |
| | C28311T | P13L | ORF9 | Yes[5] | - |
| | C25584C | - | ORF3a | No[5] | Unique common mutation in Omicron (C25584T)[5] |
| | G21987G | - | ORF2 (S) | No[5] | Unique common mutation in Omicron (Δ21989–21991 ΔGTT: Δ143V)[5] |
| | G22813G | - | ORF2 (S) | No[5] | Beta (G22813T: K417N) [4] |

# 2. SHAP value in misclassified sequences in 2-layer CNN model

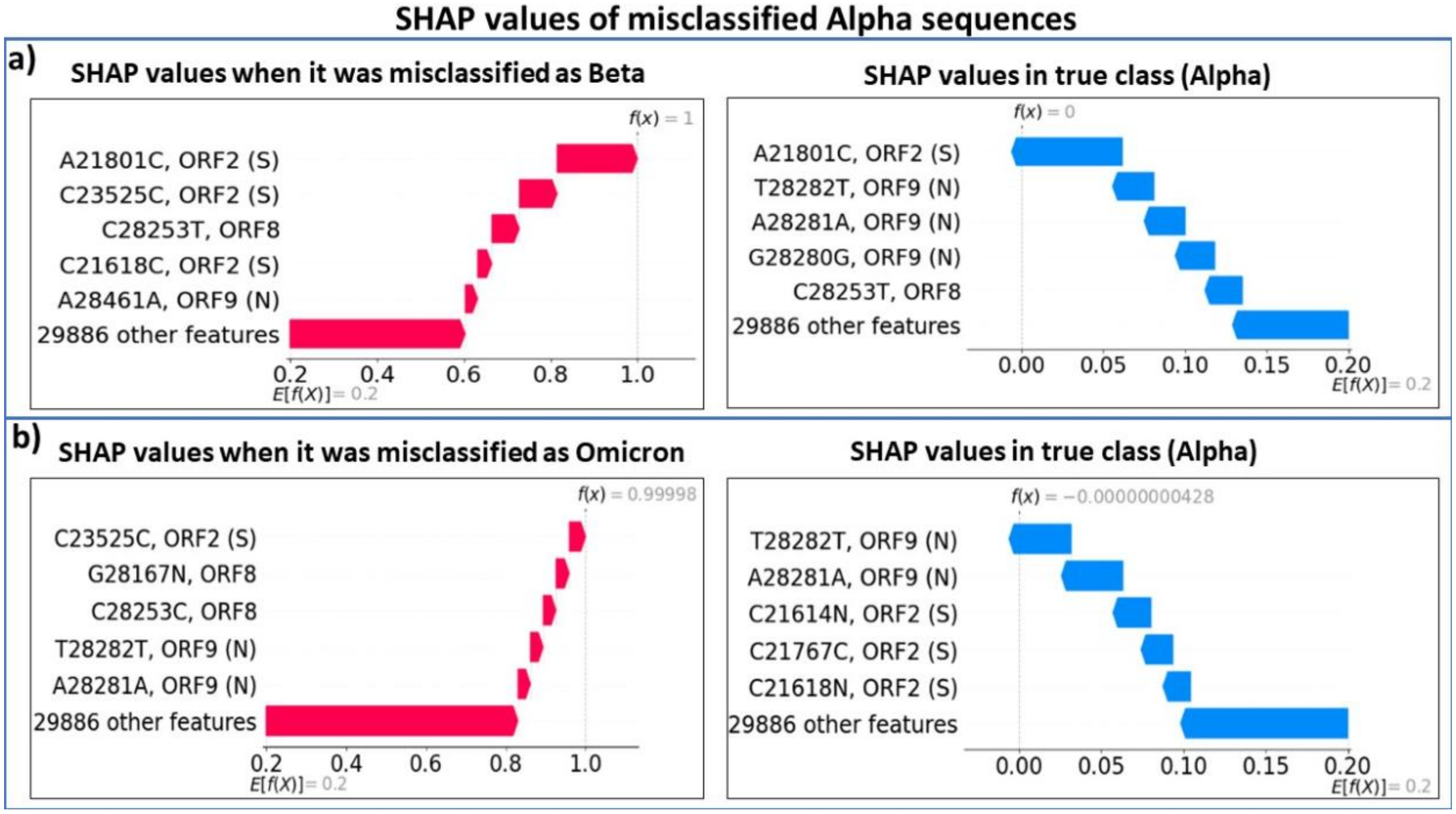

Figure S2. SHAP Analysis of Misclassified Alpha Sequences: This figure elucidates the SHAP values for two Alpha variant sequences incorrectly classified as Beta (a) and Omicron (b), identified by GISAID dataset IDs EPI_ISL_2110507 and EPI_ISL_2391826, respectively. a) Detailed examination revealed that the hallmark mutations G28280C, A28281T, and T28282A associated with the Alpha variant (D3L: ORF9) were absent in these samples and these have negative effect on model to classify the sequence as Alpha (Fig.3.b). The sequence misclassified as Beta exhibited G28280G, A28281A, T28282T, and contained additional mutations such as A21801C (D80A: ORF2 (S)) and C28253T (F120F: ORF8), which are critical for Beta classification. This misclassification was not attributed to the Gamma or Delta variants due to the absence of mutations at positions 28167, 23525, 21974 (Gamma) and 21618, 28461 (Delta), nor Omicron, as it lacked mutations at 23525, 28271, 28311. The presence of mutations A21801C and C28253T influenced the model's classification, erroneously indicating Beta rather than other variants. b) the sequence misclassified as Omicron possessed the C23604A (P681H: ORF2 (S)) mutation, shared by both Alpha and Omicron and lacked concurrent mutations with other variants, except for A23403G (D614G: ORF2 (S)), which is widespread across all VOCs. This sequence was not mistaken for Beta, Gamma, or Delta due to the non-correspondence with critical mutations at the respective positions for these variants. Notably, the base value for Omicron variant is higher than the base value of other variants, when there is no clear clue for the classification, our model thinks it belongs to Omicron.

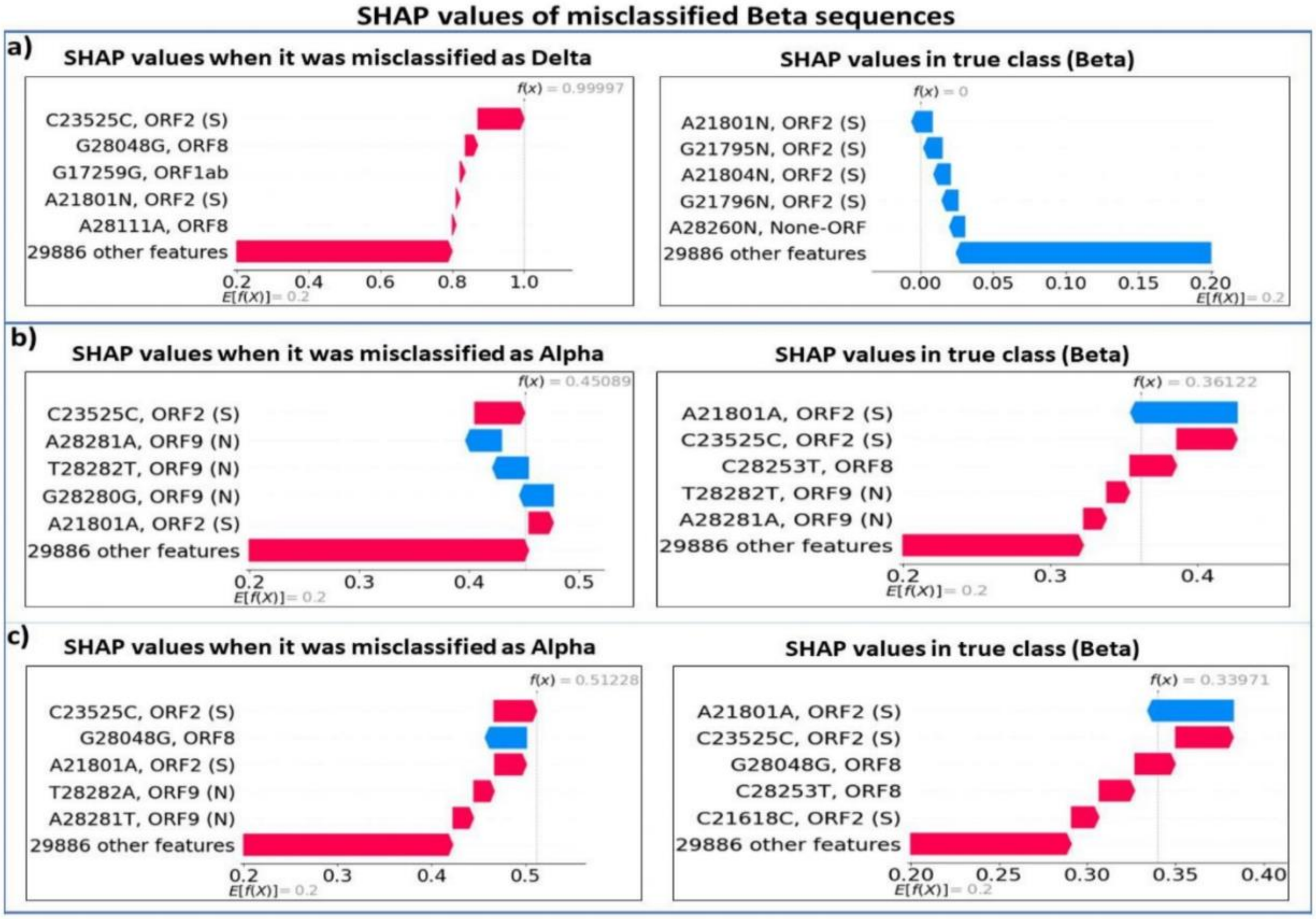

Figure S3. SHAP Analysis of Misclassified Beta Variant Sequences: Among misclassified sequences, three Beta sequences were misclassified. The key mutations associated with the Beta variant in our predictive model are A21801C (D80A: ORF2 (S)) and A22206G (D215G: ORF2 (S)). a) a Beta sequence with GISAID ID EPI_ISL_2321170 was misclassified as the Delta variant due to the presence of mutations A21801N, A22206N, and the absence of mutations at positions 28048 (associated with Alpha), 23525 and 28167, 17259 (associated with Gamma), and 23525, 28271 (associated with Omicron). The model was influenced by the presence of "C" at position 23525 and "N" at 21801, both of which are indicative of the Delta variant according to Fig.3h. b) a beta sequence with GISAID ID EPI_ISL_4742249 was misclassified as Alpha. It has no mutations at the key Beta positions. Instead, it has a C23525C, an A21801A variant, and a G28048T mutations, which are significant for the Alpha variant. This combination of mutations led the model to classify the Beta sequence as Alpha. This sequence was not identified as Gamma due to the lack of mutations at positions 23525, 28167; not as Delta due to the absence of mutations at 28461; and not as Omicron due to the absence of mutations at 23252, 28271, and 28311. c) a beta sequence with GISAID ID EPI_ISL_4742165 was incorrectly classified as Alpha. This sequence does not have mutations at position 21801 but has mutations corresponding to Alpha's critical mutations, including A28281T, T28282A, and C23252C. The absence of mutations at positions 28167 and 17259 ruled out the Gamma variant, the absence of mutations at 21618, 28461 excluded Delta, and the absence of mutations at 23252 along with a gap at 28271 (A28171_) precluded Omicron classification. These factors contributed to the incorrect classification of the Beta sequence as Alpha.

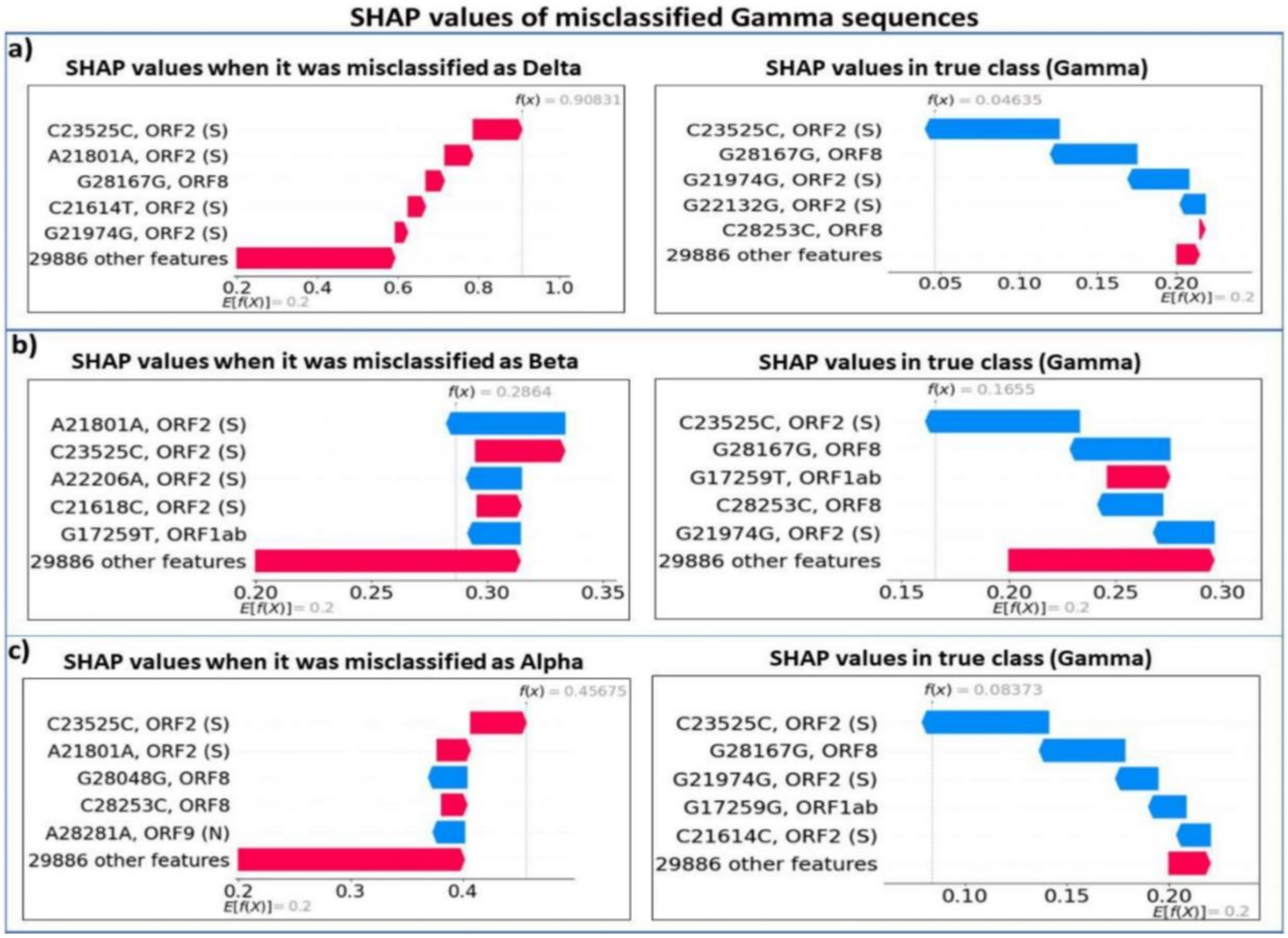

Figure S4. Analysis of misclassified Gamma sequences: Presented here are SHAP values for three Gamma variant sequences that were misclassified by the model as Delta, Beta, and Alpha variants. Notable mutations for the Gamma variant are G28167A (E92K: ORF8), G21974T (D138Y: ORF2 (S)), C23525T (H655Y: ORF2 (S)), and G17259T (E1264D) according to our CNN model. a) The Gamma sequence with GISAID ID EPI_ISL_3414795, lacking mutations at positions 28167 and 23525, was incorrectly identified as the Delta variant. The absence of these mutations negatively influenced the Gamma classification (Fig.2.f). The presence of "C" at position 23252, "G" at position 28167, and "A" at position 21801 swayed the model towards a Delta prediction. The sequence was not classified as Alpha due to a lack of mutations at 28048, 28280, 28281, 28282, nor as Beta because it had no mutations at 21801, 28252, 22206. The absence of mutations at 23252, 28271, 28311 precluded an Omicron classification. b) For the Gamma sequence with GISAID ID EPI_ISL_5801767, which the model misclassified as Beta, no mutations were found at 28167, 21974, and 23525. With no mutations at 23252, 21801 (Important to all variants in the model), 28048, and 28281 (Alpha-associated), 22206, and 28253 (Beta-associated), 21618, and 21461 (Delta-associated), the model's function outputs were closely matched. However, the combination of mutations at 23252 and 21618 outweighed the lack of mutations at 21801, 22206, and 17259, leading the model to misclassify the sequence as Beta. The sequence was not classified as Omicron due to the absence of mutations at 23252, 28271, 28311, strengthening the case against Omicron classification. c) The Gamma sequence with GISAID ID EPI_ISL_2796508, misclassified as Alpha, lacked mutations at 23252, 28167, 21974, 17259. Despite having no mutations in common with Alpha, the absence of mutations at 21801, 28253 (key for Beta), 21618, and 28461 (key for Delta), and 23252, 28271, and 28311 (key for Omicron) influenced the SHAP value summation, erroneously nudging the model towards an Alpha classification.

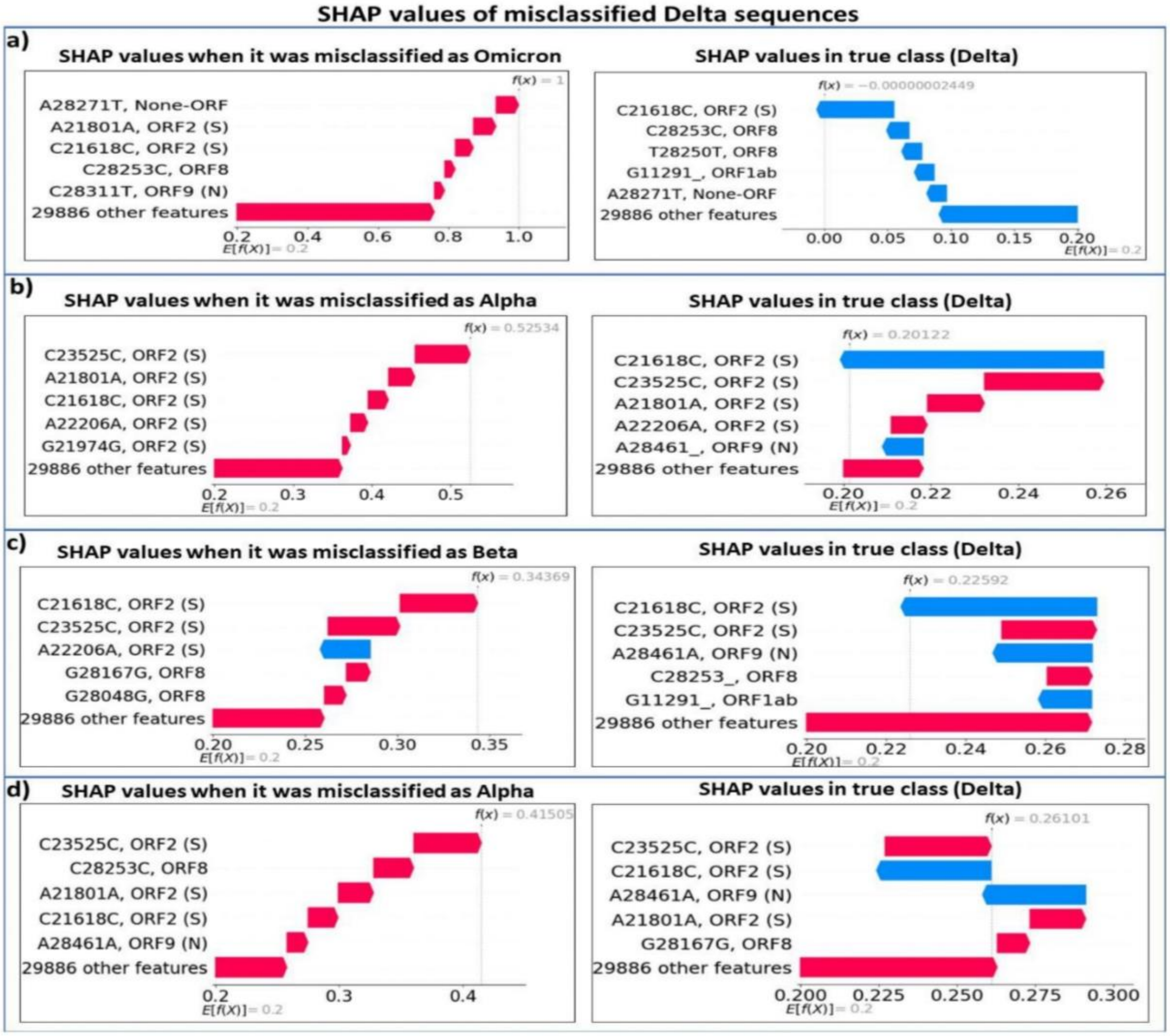

Figure S5. Analysis of misclassified Delta sequences: This figure showcases SHAP values for four Delta variant sequences that were misclassified as Omicron, Alpha, Beta, and Alpha, respectively. Key mutations that characterize the Delta variant in our models are C21618G (T19R: ORF2 (S)) and A28461G (D63G: ORF9 (N)). a) The Delta sequence with GISAID ID EPI_ISL_8769787, which is misclassified as Omicron, lacked the mutation at position 21618. It is not classified as Alpha due to the absence of mutations at positions 28048, 28280, 28281, and 28282; not as Beta because it lacked mutations at 21801 and 28253; not as Gamma due to no mutations at 23252, 28167, 21974, 17259. The presence of mutation A28271T, significant for Omicron, led the model to misclassify the sequence. b) The Delta sequence with GISAID ID EPI_ISL_8191385 is incorrectly identified as Alpha. The lack of a mutation at 21618 outweighed the importance of "no-mutation" positions 23252 (associated with Gamma and Omicron), and 21801 and 22206 (associated with Beta). This demonstrates how a deficit in critical mutations can overpower the influence of other positions. The aggregate SHAP values influenced the model to associate the sequence with Alpha, with C23252C being a pivotal factor. c) The Delta sequence with GISAID ID EPI_ISL_4233463 was misclassified as Beta. Again, the absence of a mutation at 21618 hindered correct classification. It was not identified as Alpha because of no mutations at 28048, 28280, 28281, and the combination of positions 23252 and 21618 do not sufficiently sway the model towards Delta. It was not classified as Gamma due to the lack of mutations at 23252, 28167, 21974, and while 22206 exerted some influence, it was not enough to prevent misclassification as Beta. Omicron classification was excluded due to the absence of mutations at 23252 and 28271. d) The Delta sequence with GISAID ID EPI_ISL_6068501 was also misclassified as Alpha. The missing mutation at 21618, the absence of Beta-associated mutations at 21801, 22206, and 28253, along with the presence of mutations at 23525 and 21618, pulled the model away from Delta classification. The sequence's lack of mutations at 23525, 28167, and 21974 (key for Gamma) and the absence of critical Omicron-associated mutations at 23252, 28271, 28311 ultimately led the model towards an Alpha misclassification.

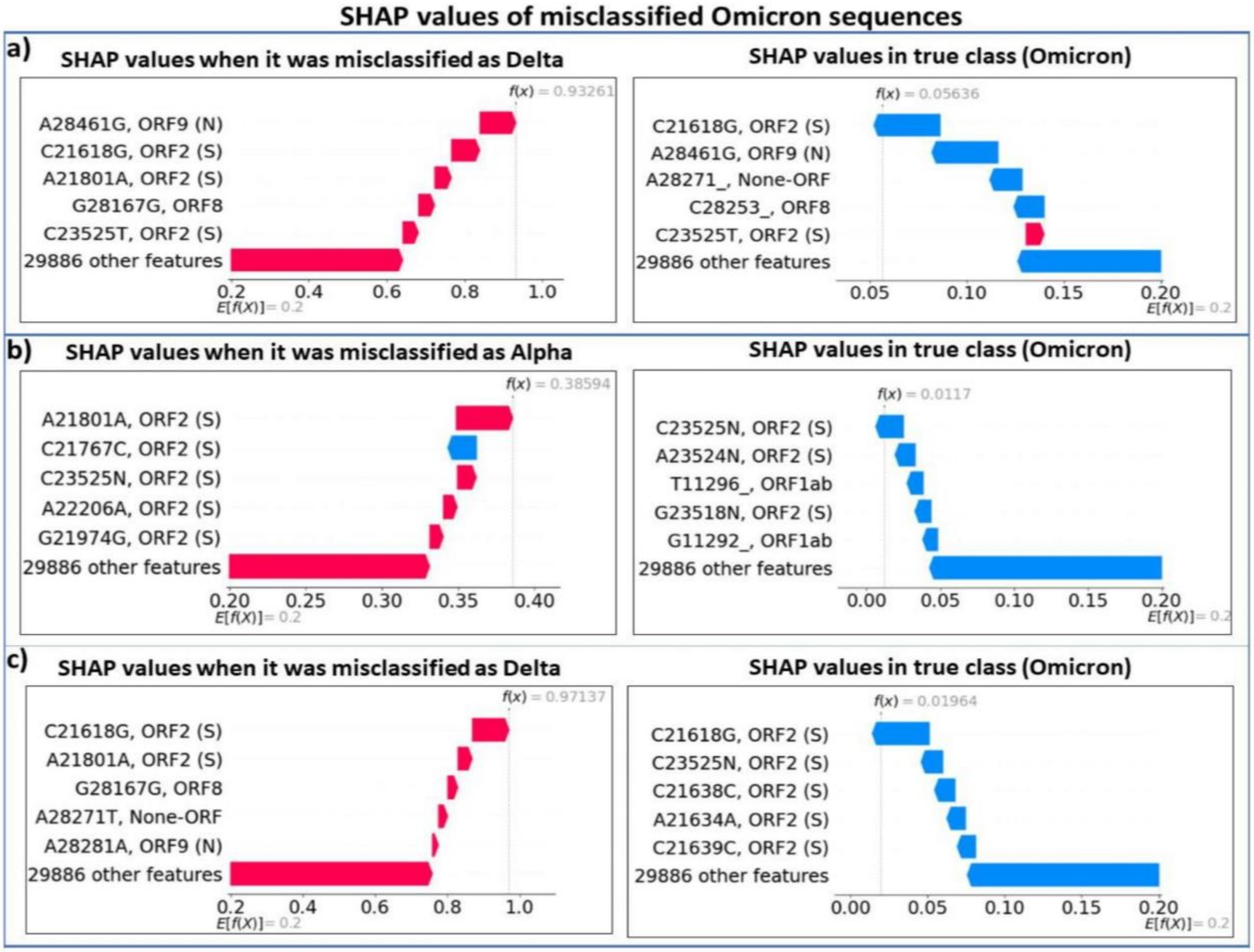

Figure S6. Analysis of misclassified Omicron sequences: This figure demonstrates SHAP values for three Omicron variant sequences incorrectly classified as Delta, Alpha, and Delta. From the model's perspective, notable Omicron mutations include C23525T (H655Y: ORF2 (S)), A28271T (between ORF8 and ORF9 (N)), and C28311T (P13L: ORF9 (N)). a) The Omicron sequence with GISAID ID EPI_ISL_9147201 exhibited a gap at A28271_, influencing the model's inability to classify it as Omicron. The mutation C21618G, a key indicator for Delta, was present, propelling the model to misclassify the sequence as Delta. The sequence could not be classified as Alpha due to the mutation C23525T (characteristic of both Omicron and Gamma) and the lack of key Alpha-associated mutations at 28048 and 28281. The absence of mutation at 21801, important for Beta, and the gap at C28253_, where C28253T is critical for Beta, exclude Beta as prediction for the sequence. Also, it does not go toward Gamma classification due to the absence of mutations at 28167 and 17259. The Delta misclassification was further cemented by the presence of Delta-significant mutations A28461G (D63G: ORF9 (N)) and A21618G (T19R: ORF2 (S)). b) The Omicron sequence with GISAID ID EPI_ISL_10205790 displayed mutation C23525N, which misled the model regarding its classification. The absence of mutations at positions 21801 and 22206 distanced the model from Beta classification, while the lack of mutations at 21974 and 17259 alongside C23525N shifted the model away from Gamma. The key point of confusion arose from the mutation C23525N, which has a negative impact in Delta but a positive albeit varied effect in Alpha as different nucleotide (Fig.3b and Fig.3h). Given this context, the model inaccurately favored Alpha for this sequence. c) Another Omicron sequence with GISAID ID EPI_ISL_8769787, carrying C23525N instead of C23525T, perplexed the model, preventing correct Omicron classification. It was not recognized as Alpha due to the lack of mutations at 28280, 28281, and 28282; not as Beta because of the absence of mutation at 21801; and not as Gamma because of no mutations at 28167, 21974, and 17259. However, the Delta-classifying mutation C21618G (T19R) led to the sequence being incorrectly classified as Delta.

## 3. Venn Diagram Between GWAS and SHAP values of VOCS

Based on Fig.5, our GWAS analysis identified 1,073 positions (excluding ORF1a due to its complete overlap with ORF1ab) where the normalized −log(p-value) is equal to 1, indicating that these positions play significant roles from a GWAS perspective.

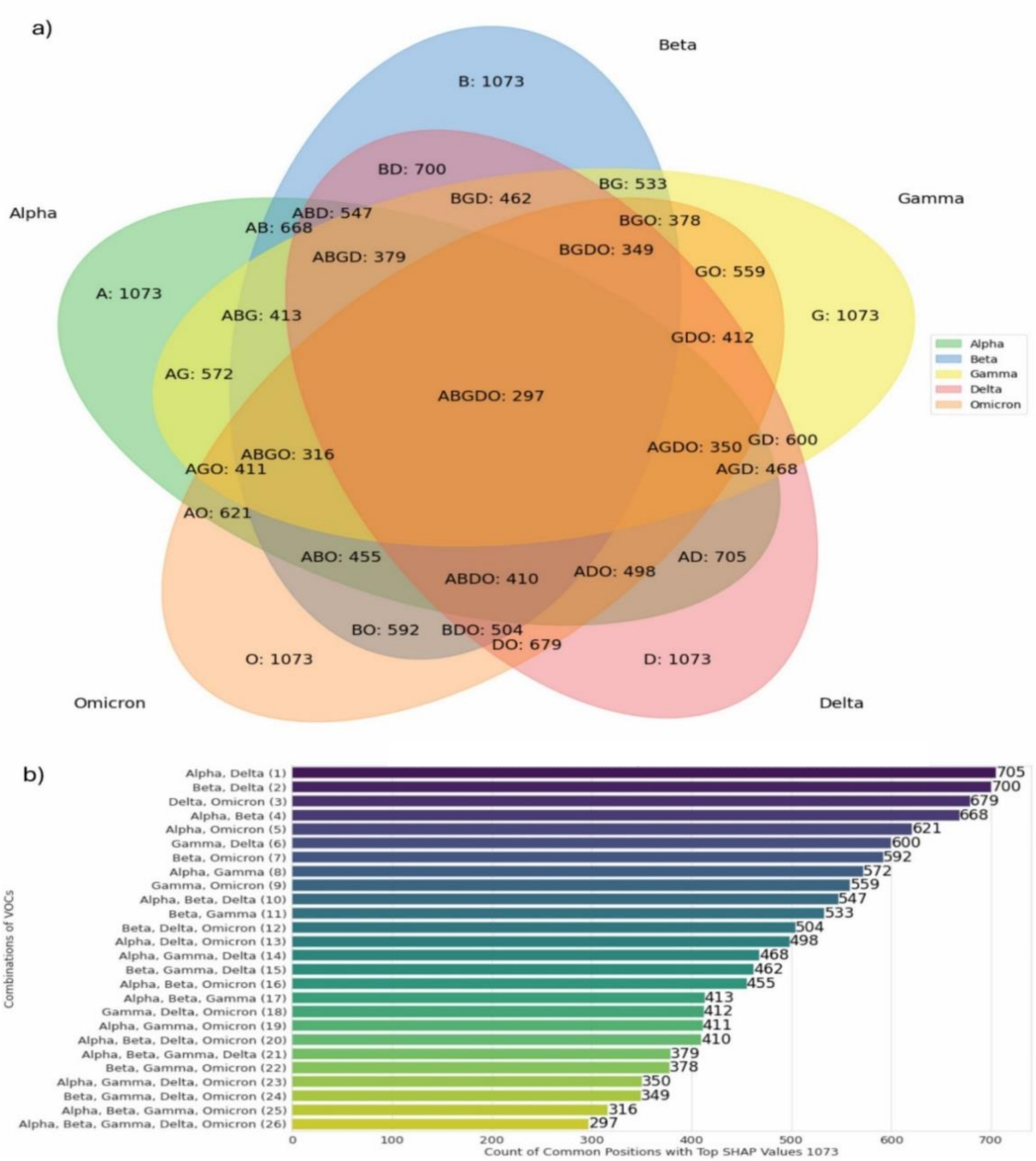

Figure S7. a) Venn diagram illustrating the overlap of the top 1,073 SHAP values across the Alpha, Beta, Gamma, Delta, and Omicron. b) Bar chart showing the number of common positions with top SHAP values across different combinations of VOCs. The categories from the Venn diagram are sorted in descending order for better clarity.

In parallel, Figure S7a presents the top 1,073 SHAP values across the VOCs, illustrating how many of these top SHAP values are shared across different combinations of VOCs. Notably, there are 297 common positions present across all VOCs. As shown in Fig. S.8, among the 297 common positions identified across all VOCs (Variants of Concern), approximately 77% are located in the spike (S) region, indicating that our model identifies a higher concentration of shared positions in the spike region across all VOCs. Additionally, we identified 145 overlapping positions between these 297 common positions and the top 1,073 positions from the GWAS analysis. Of these 145 overlapping positions, approximately 66% also belong to the spike region, further highlighting the importance of this region in SHAP.

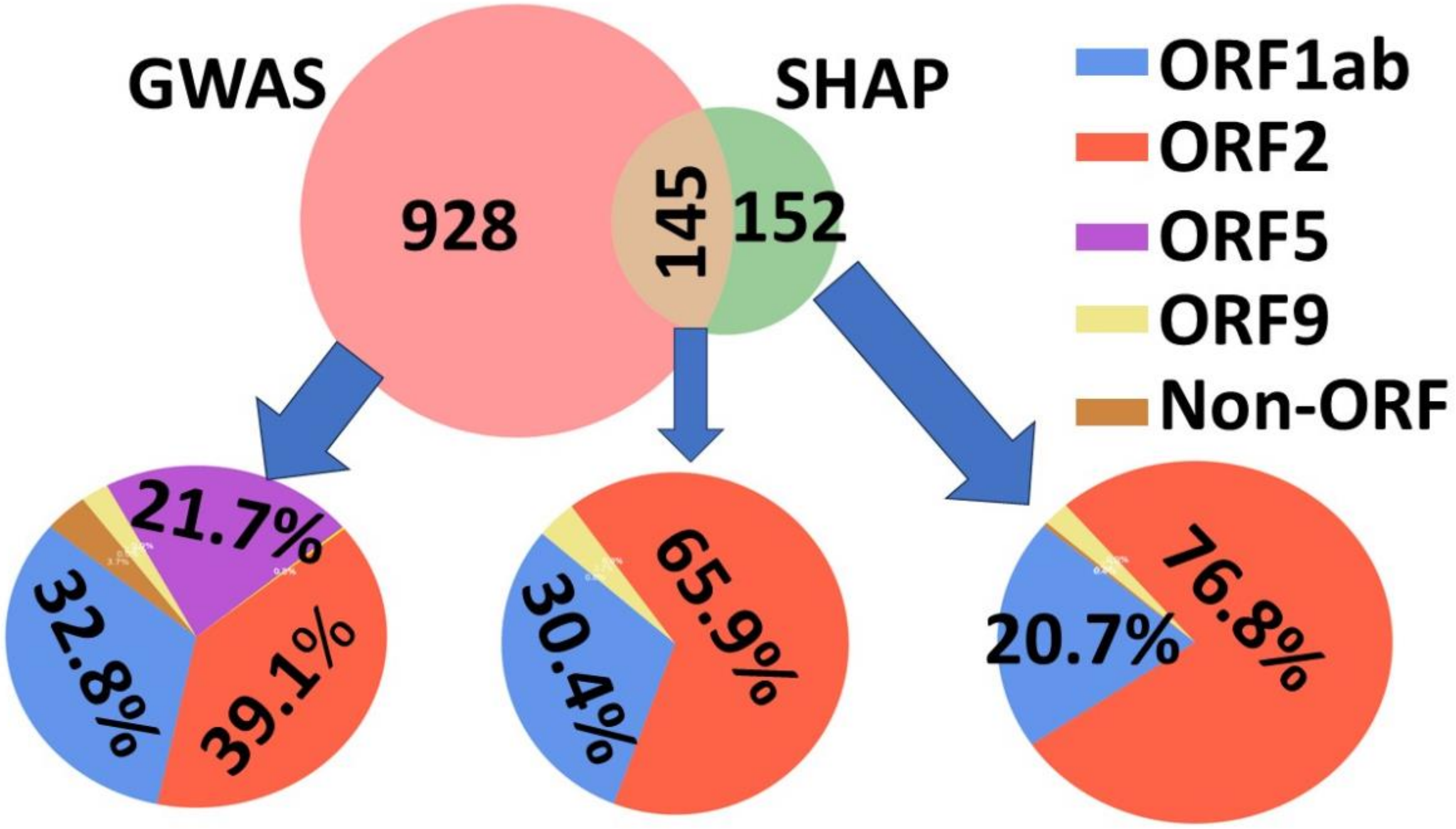

Figure S8. Comparison of key positions identified by GWAS and SHAP analyses across all Variants of Concern (VOCs). The Venn diagram shows 928 positions identified by GWAS, 152 by SHAP, and 145 overlapping positions between both methods. Pie charts illustrate the distribution of these positions across genomic regions, with a significant proportion located in the spike (S) region: 77% of the 297 common VOC positions, 66% of the 145 shared GWAS-SHAP positions, and 39.1% of the total GWAS-identified positions.

## 4. Comparison of Phylogeny and Clustering results based on SHAP values

A phylogenetic tree represents the hypothesized evolutionary relationships between organisms [7]. Fig. S9a shows the phylogenetic tree for 15 genetic sequences of VOCs, with three randomly selected sequences from each variant. The Alpha and Gamma variants are closely related, followed by Beta, Delta, and finally, Omicron, which is the most distantly related. Fig. S9.b presents a distance matrix calculated based on the absolute differences in SHAP values between pairs of VOCs. To calculate this matrix, we used distance correlation[8] to determine the dependency between each pair of VOCs. Distance correlation values range from 0 to 1, where 0 indicates independence and 1 indicates perfect dependency. This method captures non-linear associations between the VOCs, as there is no linear relationship between them. According to this analysis, the closest variants are Alpha and Beta, followed by Delta and Omicron, with Gamma being the most distantly related. This comparison reveals a discrepancy between the model's ability to discern evolutionary relationships and the relationships depicted in the phylogenetic tree. The model's predictions do not accurately reflect the true evolutionary relationships.

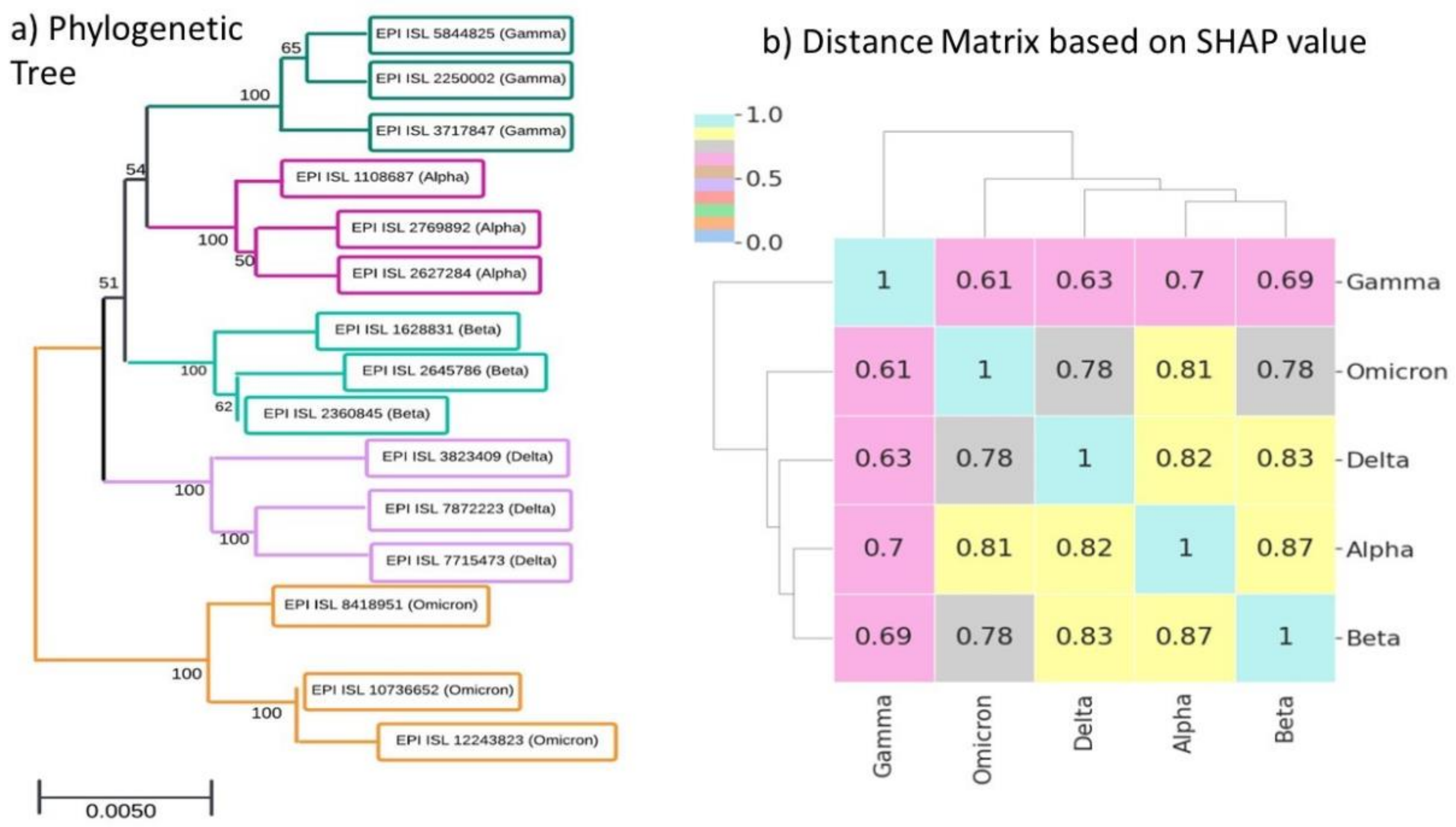

Figure S9. Comparison between Phylogeny and SHAP value Clustering for selected genomes of Variants of Concern (VOCs). a) phylogenetic tree for genetic sequences of VOCs b) Hierarchical clustering based on the correlation distance matrix between SHAP value of each VOC.